\newcounter{myctr}

\documentclass{ws-acs}
\usepackage{cite}
\usepackage{graphics}
\begin{document}

\markboth{D. Helbing, M. Sch\"onhof, H.-U. Stark, and J. A. Ho\l yst}{Emergence of Taking Turns in 
A Congestion Game} 

\catchline{}{}{}{}{}
%
%%%%%%%%%%%%%%%%%%%%%%%%%%%%%%%%%%%%%%%%%%%%%%%%%%%%%%%%%%%%%%%%%%%%

\title{HOW INDIVIDUALS LEARN TO TAKE TURNS: EMERGENCE OF ALTERNATING
COOPERATION IN A CONGESTION GAME AND THE PRISONER'S DILEMMA}

\author{\footnotesize Dirk Helbing, Martin Sch\"onhof, Hans-Ulrich Stark}

\address{Institute for Transport \& Economics, Dresden University of Technology\\
Andreas-Schubert-Str. 23, 01062 Dresden, Germany}

\author{Janusz A. Ho\l yst}

\address{Faculty of Physics  and Center of Excellence for Complex Systems Research,\\
Warsaw University of Technology, Koszykowa 75, PL-00-662 Warsaw, Poland}
\maketitle

\begin{history}
\received{(received date)}
\revised{(revised date)}
%\accepted{(Day Month Year)}
%\comby{(xxxxxxxxxx)}
\end{history}

\begin{abstract}
In many social dilemmas, individuals tend to generate a situation with low
payoffs instead of a system optimum ("tragedy of the commons"). Is the
routing of traffic a similar problem?  In order to address this question, we present experimental
results on humans playing a route choice game in a computer laboratory,
which allow one to study decision behavior in repeated games beyond the 
Prisoner's Dilemma. We will focus on whether individuals manage to find a
cooperative and fair solution compatible with the system-optimal road
usage. We find that individuals tend towards a user equilibrium with equal
travel times in the beginning. However, after many iterations, they often
establish a coherent oscillatory behavior, as taking turns performs
better than applying pure or mixed strategies. The resulting behavior is
fair and compatible with system-optimal road usage. In spite of the
complex dynamics leading to coordinated oscillations, we have identified
mathematical relationships quantifying the observed transition process.
Our main experimental discoveries for 2- and 4-person games can be
explained with a novel reinforcement learning model for an arbitrary number
of persons, which is based on past experience and trial-and-error
behavior. Gains in the average payoff seem to be an important
driving force for the innovation of time-dependent response patterns, 
i.e. the evolution of more complex strategies.
Our findings are relevant for decision support systems and
routing in traffic or data networks.
\end{abstract}


\section{Introduction} \label{Sec0}

Congestion is a burden of today's traffic systems, affecting the economic prosperity
of modern societies. Yet, the optimal distribution of vehicles over alternative routes 
is still a challenging problem and uses scarce resources (street capacity) in an inefficient way.
Route choice is based on interactive, but decentralized individual decisions, which cannot be well 
described by classical utility-based decision models \cite{NJP}. Similar
to the minority game \cite{minority,m2,m3}, 
it is reasonable for different people to react to the same situation or information
in {\em different} ways. As a consequence, individuals tend to develop characteristic
response patterns or roles \cite{bonn}. Thanks to this differentiation process, individuals learn to coordinate 
better in the course of time. However, according to current knowledge, selfish routing 
does not establish the system optimum of minimum overall travel times.
It rather tends to establish the Wardrop equilibrium, a special user or Nash equilibrium
characterized by equal travel times on all alternative
routes chosen from a certain origin to a given destination (while routes with longer travel
times are not taken) \cite{Wardrop}. 
\par
Since Pigou\cite{Pigou},  it has been suggested to resolve the problem of inefficient road usage by congestion charges,
but are they needed? Is the missing establishment of a sytem optimum just a problem of varying
traffic conditions and changing origin-destination pairs, which make route-choice decisions
comparable to one-shot games? Or would individuals in an {\em iterated} setting of a day-to-day route 
choice game with identical conditions spontaneously establish cooperation in order to increase their
returns, as the folk theorem suggests \cite{folktheorem}? 
\par
How would such a cooperation look like?
Taking turns could be a suitable solution \cite{Schelling}. While simple symmetrical cooperation is 
typically found for the repeated Prisoner's Dilemma 
\cite{AxeHa81,AxeDi88,kinship,reciprocity,similarity,reputation,loners,finite,Schweitzer,spatial,volunteering,variation}, 
emergent alternating reciprocity has been recently discovered for the games Leader and Battle of the 
Sexes \cite{BrColman}.\footnote{See Fig.~\ref{fig2} for a specification of these games.} 
Note that such coherent oscillations are a time-dependent, but deterministic form of 
individual decision behavior, which can establish a persistent phase-coordination, 
while mixed strategies, i.e. statistically varying decisions, can establish cooperation only by
chance or in the statistical average. This difference is particularly important when the
number of interacting persons is small, as in the particular route choice game discussed below.
\par
Note that oscillatory behavior has been found in iterated games before:
\begin{itemize}
\item In the rock-paper-scissors game \cite{volunteering}, 
cycles are predicted by the game-dynamical equations due to unstable stationary solutions \cite{Hofbauer}.
\item Oscillations can also result by coordination problems \cite{Arthur,Huberman,Huberman2,Namatame},
at the cost of reduced system performance.
\item Moreover, blinker strategies may survive in repeated games played by a mixture of finite automata \cite{Binmore}
or result through evolutionary strategies \cite{BrColman,minority,m2,m3,m4,m5,m6,m7}. 
\end{itemize}
However, these oscillation-generating mechanisms are clearly to be distinguished from
the establishment of phase-coordinated alternating reciprocity we are interested in (coherent 
oscillatory cooperation to reach the system optimum).
\par
Our paper is organized as follows: In Section~\ref{Sec1}, we will formally introduce the
route choice game for $N$ players, including issues like the Wardrop equilibrium \cite{Wardrop} and
the Braess paradox \cite{Braess}. Section~\ref{Sec2} will focus on the special case of
the 2-person route choice game, compare it with the minority game \cite{Arthur,minority,m2,m3,m4,m5,m6,m7}, 
and discuss its place in the classification scheme of symmetrical 2x2 games. 
This section will also reveal some apparent shortcomings of the previous
game-theoretical literature:
\begin{itemize}
\item While it is commonly stated that among the 12 ordinally distinct, symmetrical 2x2 games
\cite{Rapoport,BrColman} only 4 archetypical 2x2 games
describe a strategical conflict (the Prisoner's Dilemma, the Battle of the Sexes, Chicken, and Leader)
\cite{Rapoport2,BrColman,ColmanBook}, we will show that, 
for specific payoffs, the route choice game (besides Deadlock) also represents an interesting strategical conflict, 
at least for iterated games.
\item The conclusion that conservative driver behavior is best, i.e. it does not pay off to change routes \cite{press,inSelten,ssrn},
is restricted to the special case of route-choice games with a system-optimal user equilibrium. 
\item It is only half the truth that cooperation in the iterated Prisoner's Dilemma is characterized by symmetrical
behavior \cite{BrColman}. Phase-coordinated asymmetric reciprocity is possible as well, as in some other
symmetrical 2x2 games \cite{BrColman}. 
\end{itemize}
New perspectives arise by less restricted specifications of the payoff values.
\par
In section~\ref{Sec3}, we will discuss empirical results of laboratory experiments with humans 
\cite{ColmanBook,CamererBook,beheco}. According to these, reaching a phase-coordinated alternating state 
is only one problem. Exploratory behavior and suitable punishment strategies are important 
to establish asymmetric oscillatory reciprocity as well \cite{BrColman,Crowley}. Moreover,
we will discuss several coefficients characterizing individual behavior and 
chances for the establishment of cooperation. In section~\ref{Sec4}, we will present multi-agent computer simulations
of our observations, based on a novel win-stay, lose-shift \cite{NovakSigmund,Posch} strategy, which is a special
kind of reinforcement learning strategy \cite{Flache}. This approach is based
on individual historical experience \cite{Camerer} and, thereby, clearly
differs from the selection of the best-performing strategy
in a set of hypothetical strategies as assumed in studies based on 
evolutionary or genetical algorithms \cite{Binmore,minority,m2,m3,m4,m5,BrColman}. The final section
will summarize our results and discuss their relevance for game theory and
possible applications such as data routing algorithms \cite{Korilis,Wolpert}, advanced driver information
systems \cite{bonsall,Hu,MahJou,selten,schreck,Nagel,Johnson,Yama}, or road pricing \cite{Pigou}.

\section{The Route Choice Game} \label{Sec1}

\begin{figure}[htbp]
\begin{center}
\includegraphics[width=12cm]{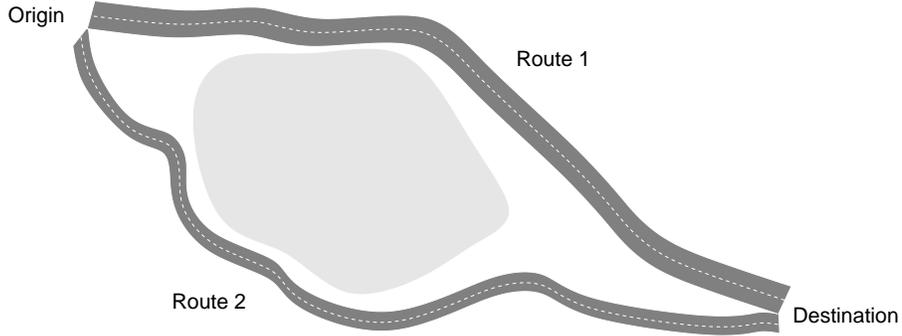} 
\end{center}
\caption[]{Illustration of the investigated day-to-day route choice scenario. We study the dynamic
decision behavior in a repeated route choice game, where a given destination can be reached from
a given origin via two different routes, a freeway (route 1) and a side road (route 2).\label{fig1}}
\end{figure}
In the following, we will investigate a scenario with two alternative routes between a certain origin
and a given destination, say, between two places or towns A and B (see Fig.~\ref{fig1}). We are interested in the case
where both routes have different capacities, say a freeway and a subordinate or side road.
While the freeway is faster when it is empty, it may be reasonable to use the side road when the
freeway is congested. 
\par
The ``success'' of taking route $i$ could be measured in terms of its
inverse travel time $1/T_i(N_i) = V_i(N_i)/L_i$, where $L_i$ is the length of route $i$ and $V_i(N_i)$
the average velocity when $N_i$ of the $N$ drivers have selected route $i$.
One may roughly approximate the average vehicle speed $V_i$ on route $i$ by the linear relationship \cite{Greenshield}
\begin{equation}
       V_i(N_i) = V_i^0 \left( 1 - \frac{N_i(t)}{N_i^{\rm max}} \right) \, ,
\end{equation}
where $V_i^0$ denotes the maximum velocity (speed limit)
and $N_i^{\rm max}$ the capacity, i.e. the maximum possible number of vehicles on route $i$.
With $A_i = V_i^0/L_i$ and $B_i = V_i^0/(N_i^{\rm max} L_i)$, the
inverse travel time then obeys the relationship
\begin{equation}
1/T(N_i) = A_i - B_i N_i \, ,
\end{equation}
which is linearly decreasing with the road occupancy $N_i$. 
Other monotonously falling relationships $V_i(N_i)$ would make the expression for the
inverse travel times non-linear, but they would probably not lead to qualitatively different conclusions. 
\par
The user equilibrium of equal travel times is found for a fraction
\begin{equation}
 \frac{N_1^{\rm e}}{N} = \frac{B_2}{B_1+B_2} + 
 \frac{1}{N} \frac{A_1-A_2}{B_1+B_2} 
\end{equation} 
of persons choosing route 1. In contrast, the system optimum corresponds to the maximum of
the overall inverse travel times $N_1/T_1(N_1) + N_2/T_2(N_2)$ and is found for 
the fraction 
\begin{equation}
 \frac{N_1^{\rm o}}{N} = \frac{B_2}{B_1+B_2} + 
 \frac{1}{2N} \frac{A_1-A_2}{B_1+B_2} 
\end{equation} 
of 1-decisions. The difference between both fractions vanishes in the limit $N\rightarrow \infty$.
Therefore, only experiments with a few players allow to find out, 
whether the test persons adapt to the user equilibrium or to the system optimum. We will see that
both cases have completely different dynamical implications: While the most successful strategy to
establish the user equilibrium is to stick to the same decision in subsequent iterations \cite{inSelten,ssrn,NJP},
the system optimum can only be reached by a time-dependent strategy (at least, if no participant is ready 
to pay for the profits of others). 
\par
Note that alternative routes can reach comparable travel times only
when the total number $N$ of vehicles is large enough to fulfil the relationships
$P_1(N) < P_2(0) = A_2$ and $P_2(N) < P_1(0) = A_1$. Our route choice game
will address this traffic regime and additionally assume $N \le N_i^{\rm max}$. 
The case $N_i = N_i^{\rm max}$ corresponds
to a complete gridlock on route $i$. 
\par
Finally, it may be interesting to connect the previous quantities with
the vehicle densities $\rho_i$ and the traffic flows $Q_i$:
If route $i$ consists of $I_i$ lanes, 
the relation with the average vehicle density is $\rho_i(N_i) = N_i/(I_iL_i)$, and the relation with
the traffic flow is $Q_i(N_i) = \rho_i V_i(N_i) = N_i / [I_i T_i(N_i)]$. 
\par
In the following, we will linearly transform the inverse travel time $1/T_i(N_i)$ in order
to define the so-called payoff
\begin{equation}
 P_i(N_i) = C_i - D_i N_i
\end{equation}
for choosing route $i$. The payoff parameters $C_i$ and $D_i$ depend on the parameters $A_i$, $B_i$, 
and $N$, but will be taken constant. We have scaled the parameters so that we
have the payoff $P_i(N_i^{\rm e}) =0$ (zero payoff points) in the user equilibrium and
the payoff $N_1 P_1(N_1^{\rm o}) + N_2 P_2(N-N_1^{\rm o}) =100 N$ (an average of
100 payoff points) in the system optimum. This serves to reach generalizable results and to provide a 
better orientation to the test persons.
\par
Note that the investigation of social (multi-person) games with linearly falling payoffs
is not new \cite{Huberman}. For example, Schelling \cite{Schelling} has discussed situations with ``conditional
externality'', where the outcome of a decision depends on the independent decisions of potentially many 
others \cite{Schelling}. Pigou has addressed this problem, which has been recently focused on by
Schreckenberg and Selten's project SURVIVE \cite{press,inSelten,ssrn} and others \cite{bonsall,reddy,MahJou}. 
\par
The route choice game is a special 
congestion game \cite{Rosenthal,Garcia,ModererShapley}. More precisely speaking, it is a 
multi-stage symmetrical $N$-person single commodity congestion game \cite{spirakis}. 
Congestion games belong to the class of ``potential games'' \cite{potgames}, 
for which many theorems are available. For example, it is known that there always exists
a Wardrop equilibrium \cite{Wardrop} with essentially unique Nash flows \cite{Beckmann}. This is
characterized by the property that no individual driver can decrease his or her travel time by a different
route choice. If there are several alternative routes from a given origin to a
given destination, the travel times on all used alternative routes in the Wardrop equilibrium is 
the same, while roads with longer travel 
times are not used. However, the Wardrop equilibrium as expected outcome of selfish routing
does not generally reach the system optimum, i.e. minimize the total travel times. Nash flows are 
often inefficient, and selfish behavior implies the possibility of decreased network performance.\footnote{For 
more details see the work by T. Roughgarden.} This is particularly 
pronounced for the Braess paradox \cite{Braess,Roughgarden2001}, according to
which additional streets may 
sometimes increase the overall travel time and reduce the throughput of a road network. 
The reason for this is the possible existence of badly performing 
Nash equilibria, in which no single person can improve his or her payoff 
by changing the decision behavior. 
\par
In fact, recent laboratory experiments indicate that, in a ``day-to-day route choice scenario'' 
based on selfish routing, the distribution of individuals over the alternative routes is 
fluctuating around the Wardrop equilibrium
\cite{selten,NJP}. Additional conclusions from the laboratory experiments 
by Schreckenberg, Selten {\em et al.} are as follows \cite{inSelten,ssrn}:
\begin{itemize}
\item Most people, who change their decision frequently, respond 
to their experience on the previous day (i.e. in the last iteration). 
\item There are only a few different behavioral patterns: direct responders (44\%),
contrarian responders (14\%), and conservative persons, who do not respond to
the previous outcome. 
\item It does not pay off to react to travel time information in a sensitive way,
as conservative test persons reach the smallest travel times (the largest payoffs) on average.
\item People's reactions to short term travel forecasts can invalidate these. Nevertheless,
travel time information helps to match the Wardrop equilibrium, so that excess travel
times due to coordination problems are reduced.
\end{itemize}
A closer experimental analysis based on longer time series (i.e. more iterations) for
smaller groups of test persons reveals a more detailed picture \cite{bonn}:
\begin{itemize}
\item Individuals do not only show an adaptive behavior to the travel times on the previous
day, but also change their response pattern in time \cite{bonn,kluegl}.
\item In the course of time, one finds a differentiation process which leads to the development
of characteristic, individual response patterns, which tend to be almost deterministic
(in contrast to mixed strategies).
\item While some test persons respond to small differences in travel times, others only
react to medium-sized deviations, further people respond to large deviations, etc. 
In this way, overreactions of the group
to deviations from the Wardrop equilibrium are considerably reduced.
\end{itemize}
Note that the differentiation of individual behaviors is a
way to resolve the coordination problem to match the Wardrop equilibrium exactly, i.e.
which participant should change his or her
decision in the next iteration in order to compensate for a deviation 
from it. This implies that the fractions of specific behavioral response patterns 
should depend on the parameters of the payoff function. 
A certain fraction of ``stayers'', who
do not respond to travel time information, can improve 
the coordination in the group, i.e. the overall performance. However, 
stayers can also prevent the establishment of a system optimum, if 
alternating reciprocity is needed, see Eq.~(\ref{coop}).

\section{Classification of Symmetrical 2x2 Games} \label{Sec2}

In contrast to previous laboratory experiments, we have studied the route choice game
not only with a very high number of repetitions, but also with a small number $N \in \{2,4\}$ of test persons,
in order to see whether the system optimum or the Wardrop equilibrium
is established. Therefore, let us shortly discuss how
the 2-person game relates to previous game-theoretical studies.
\par\begin{figure}[htbp]
\begin{center}
\includegraphics[width=10cm]{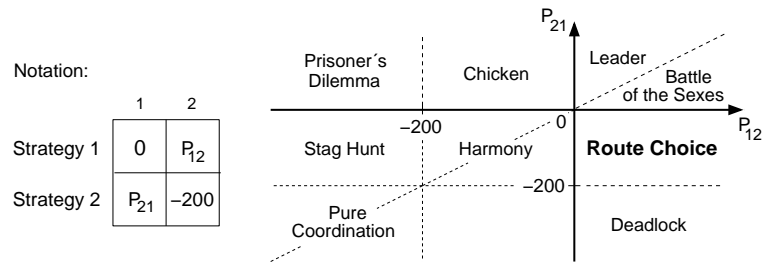} 
\end{center}
\caption[]{Classification of symmetrical 2x2 games according to their payoffs $P_{ij}$. 
Two payoff values have been kept constant as payoffs may be linearly transformed and the two strategies
of the one-shot game renumbered. Our choice of $P_{11} =0$ and $P_{22} = -200$
was made to define a payoff of 0 points in the user equilibrium and an average payoff of 100
in the system optimum of our investigated route choice game with $P_{12}=300$ and $P_{21} = -100$.\label{fig2}}
\end{figure}
\begin{figure}[htbp]
\begin{center}
\includegraphics[width=12.5cm]{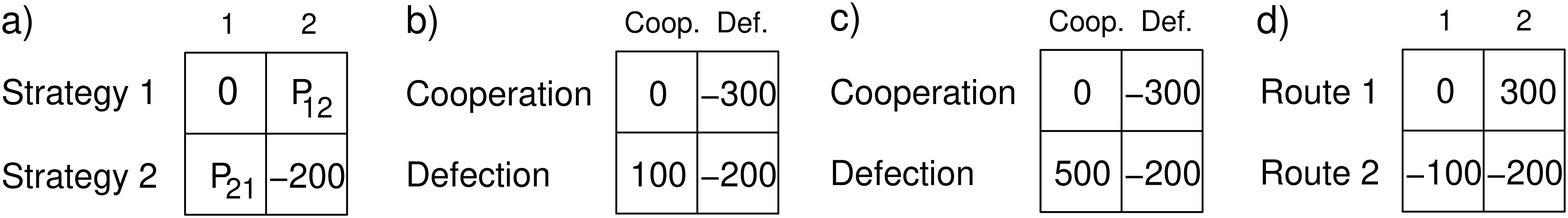} 
\end{center}
\caption[]{Payoff specifications of the symmetrical 2x2 games investigated in this paper.
a) General payoff matrix underlying the classification scheme 
of Fig.~\ref{fig2}. b), c) Two variants of the Prisoner's Dilemma. d) Route choice
game with a strategical conflict between the user equilibrium and the system optimum.\label{fig3}}
\end{figure}

Iterated symmetrical two-person games have been intensively studied \cite{ColmanBook,CamererBook}, including 
Stag Hunt, the Battle of the Sexes, or the Chicken Game (see Fig.~\ref{fig2}). They can all be
represented by a payoff matrix of the form ${\mathbf{P}}=(P_{ij})$, where 
$P_{ij}$ is the success (``payoff'') of person 1 in a one-shot game 
when choosing strategy $i\in \{1,2\}$ 
and meeting strategy $j\in \{1,2\}$. The respective payoffs of the second person are
given by the symmetrical values $P_{ji}$. Figure \ref{fig2} shows a systematics of the 
previously mentioned and other kinds of symmetrical two-person games \cite{Lindgren}. 
The relations 
\begin{equation}
P_{21} > P_{11} > P_{22} > P_{12} \, , 
\end{equation}
for example, define a Prisoner's Dilemma.
In this paper, however, we will mainly focus on the 2-person route choice game 
defined by the conditions 
\begin{equation}
 P_{12} > P_{11} > P_{21} > P_{22} 
\end{equation}
(see Fig.~\ref{fig3}).  Despite some common properties, this game 
differs from the minority game \cite{minority,m2,m3} 
or El Farol bar problem \cite{Arthur} with $P_{12}, P_{21} > P_{11}, P_{22}$,
as a minority decision for alternative 2 is less profitable than a majority decision for alternative 1.
Although oscillatory behavior has been found 
in the minority game as well \cite{minority,m2,m4,Bottazzi,minor1}, 
an interesting feature of the route choice experiments discussed in the following is the
regularity and phase-coordination (coherence) of the oscillations.
\par
The 2-person route choice game fits well into the classification scheme of symmetrical 2x2 games.
In Rapoport and Guyer's taxonomy of 2x2 games \cite{Rapoport}, the 2-person route choice game appears 
on page 211 as game number 7 together with four other games with strongly stable equilibria.
Since then, the game has almost been forgotten and did not have a commonly known interpretation or name.
Therefore, we suggest to name it the 2-person ``route choice game''. Its place in the 
extended Eriksson-Lindgren scheme of symmetrical 2x2 games is graphically illustrated
in Fig.~\ref{fig2}. 
\par
According to the game-theoretical literature, there are 
12 ordinally distinct, symmetric 2x2 games \cite{Rapoport}, 
but after excluding strategically trivial games in the sense of having equilibrium
points that are uniquely Pareto-efficient, there remain four archetypical 2x2 games:
the Prisoner's Dilemma, the Battle of the Sexes, Chicken (Hawk-Dove), and Leader \cite{Rapoport2}. 
However, this conclusion is only correct, if the four payoff values $P_{ij}$ are specified by the
four values $\{1,2,3,4\}$. Taking different values would lead to a different conclusion: 
If we name subscripts so that $P_{11} > P_{22}$, 
a strategical conflict between a user equilibrium and the system optimum 
results when 
\begin{equation}
P_{12}+P_{21} > 2P_{11} \, . 
\label{condition}
\end{equation}
{\em Our conjecture is that players tend to develop alternating forms of reciprocity if this
condition is fulfilled, while symmetric reciprocity is found otherwise.} This has the following
implications (see Fig.~\ref{fig2}):
\begin{itemize}
\item If the 2x2 games Stag Hunt, Harmony, or Pure Coordination 
are repeated frequently enough, we expect always a symmetrical form of cooperation.
\item For Leader and the Battle of the Sexes, we expect the establishment of asymmetric reciprocity,
as has been found by Browning and Colman with a computer simulation 
based on a genetic algorithm incorporating mutation and crossing-over \cite{BrColman}.
\item For the games Route Choice, Deadlock, Chicken, and Prisoner's Dilemma
both, symmetric (simultaneous) and asymmetric (alternating) forms of cooperation are possible, depending 
on whether condition (\ref{condition}) is fulfilled or not. Note that this condition
cannot be met for some games, if one restricts to ordinal payoff 
values $P_{ij} \in \{1,2,3,4\}$ only. Therefore, this interesting problem has
been largely neglected in the past (with a few exceptions, e.g. \cite{alter}).  
In particular, convincing experimental evidence of alternating reciprocity is missing.
The following sections of this paper will, therefore, not only propose a simulation model, but 
also focus on an experimental study of this problem, which promises interesting new results.
\end{itemize}

\section{Experimental Results} \label{Sec3}

\begin{figure}[htbp]
\begin{center}
\includegraphics[width=13cm]{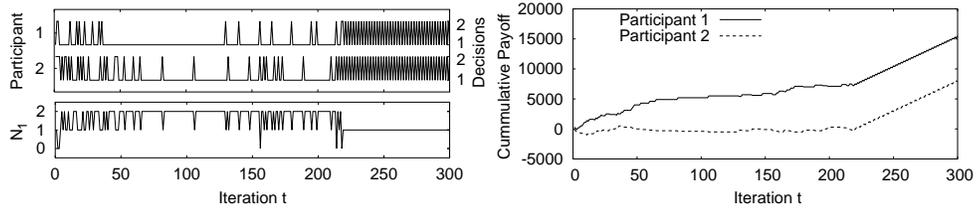} 
\end{center}
\caption[]{Representative example for the emergence of coherent oscillations 
in a 2-person route choice experiment with the parameters specified in Fig.~\ref{fig3}d.
Top left: Decisions of both participants over 300 iterations. Bottom left: Number $N_1(t)$ of 1-decisions
over time $t$. Note that $N_1 = 1$ corresponds to the system optimum, while
$N_1=2$ corresponds to the user equilibrium of the one-shot game. Right:
Cumulative payoff of both players in the course of time $t$ (i.e. as a function of the number of
iterations). Once the coherent oscillatory cooperation is established ($t>220$), both individuals have 
high payoff gains on average.\label{fig4}}
\end{figure}
\begin{figure}[htbp]
\begin{center}
\includegraphics[width=13cm]{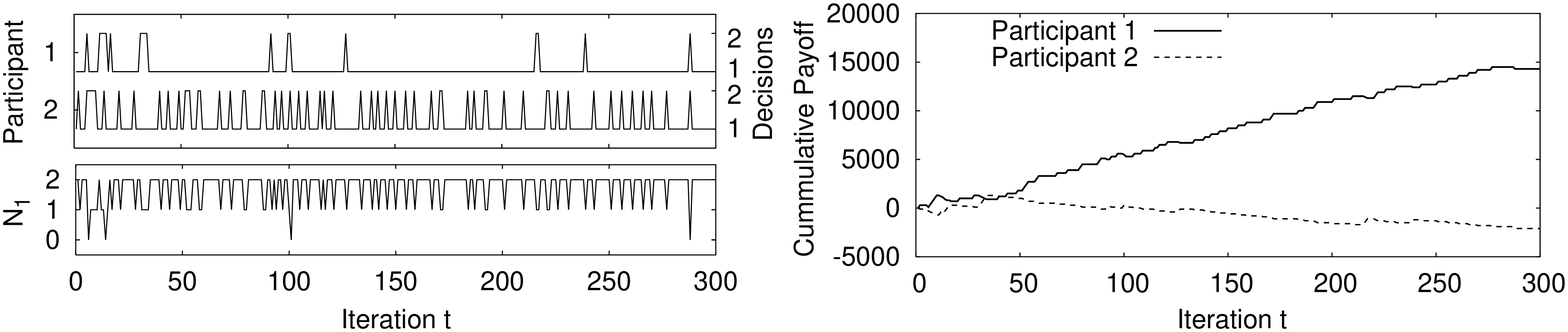} 
\end{center}
\caption[]{Representative example for a 2-person route choice experiment, in which no
alternating cooperation was established. Due to the small changing frequency of participant 1, there were not enough
cooperative episodes that could have initiated coherent oscillations.
Top left: Decisions of both participants over 300 iterations. Bottom left: Number $N_1(t)$ of 1-decisions
over time $t$. Right: The cumulative payoff of both players in the course of time $t$ shows that
the individual with the smaller changing frequency has higher profits.\label{fig5}}
\end{figure}
\begin{figure}[htbp]
\begin{center}
\includegraphics[width=6cm]{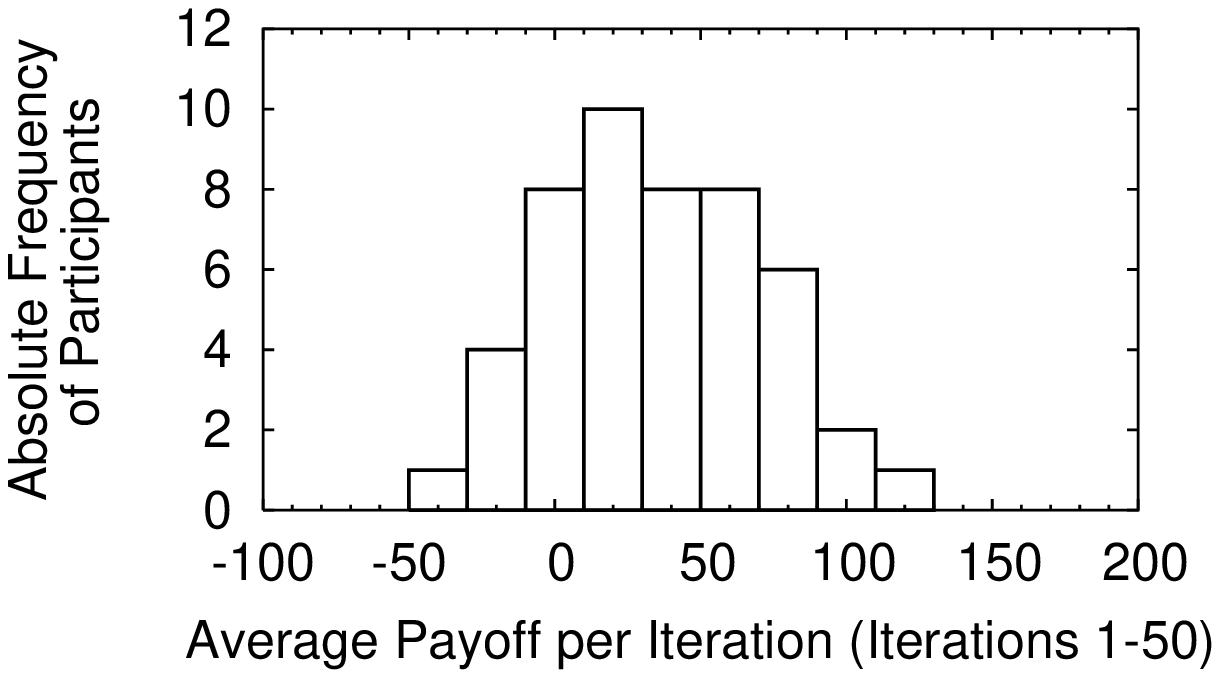} 
\includegraphics[width=6cm]{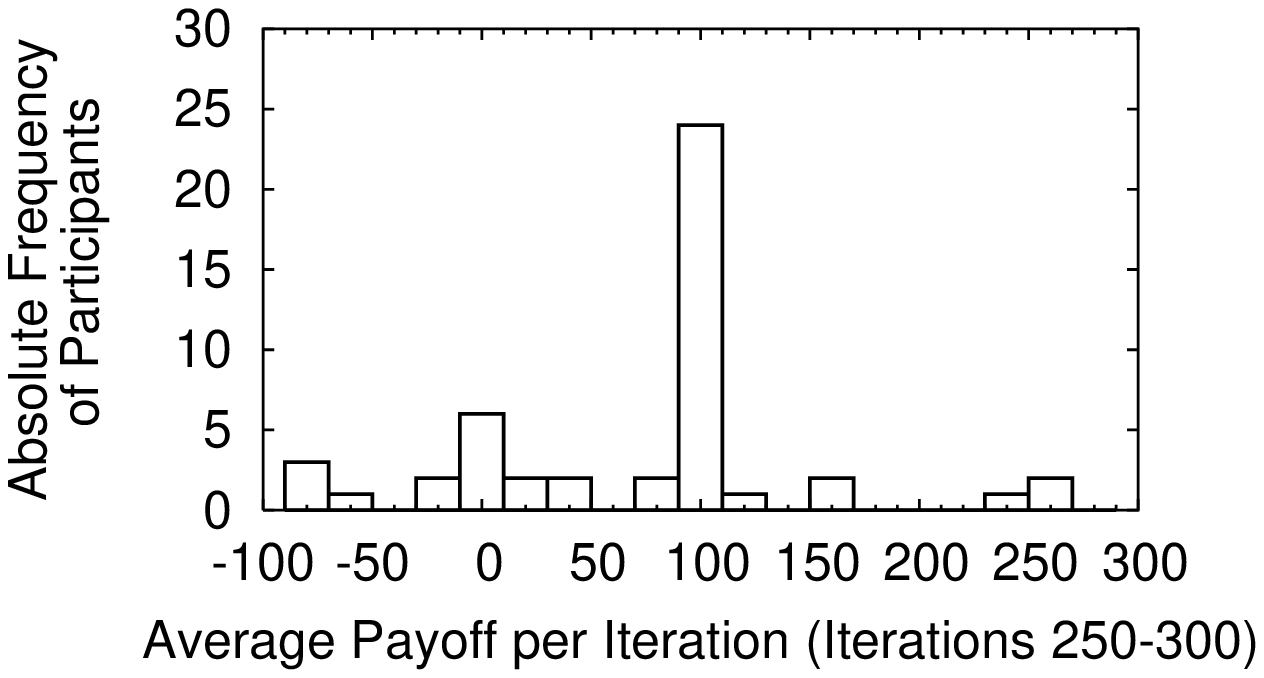} 
\end{center}
\caption[]{Frequency distributions of the average payoffs of the 48 players participating in our
24 two-person route choice experiments. Left: Distribution during the first 50 iterations. Right:
Distribution between iterations 250 and 300.  
The initial distribution with a maximum close to 0 points (left) indicates a tendency
towards the user equilibrium corresponding to the dominant strategy of the one-shot game.
However, after many iterations, many individuals learn to establish the system optimum with a payoff 
of 100 points (right).\label{martin}}
\end{figure}

Altogether we have carried out more than 80 route choice experiments with different 
experimental setups, all with different participants. In the 24 two-person [12 four-person] 
experiments evaluated here (see Figs.~\ref{fig4}--\ref{phase}), test persons were instructed to 
choose between two possible routes between the same origin and destination.
They knew that route 1 corresponds to a `freeway' (which may be fast or congested),
while route 2 represents an alternative route (a `side road'). Test persons were also informed that, 
if two [three] participants would choose route 1, everyone would receive 0 points, while   
if half of the participants would choose route 1, they would receive the maximum average amount 
of 100 points, but 1-choosers would profit at the cost of 2-choosers. Finally, participants were told that
everyone could reach an average
of 100 points per round with variable, situation-dependent decisions, and that the
(additional) individual payment after the experiment would depend on their cumulative 
payoff points reached in at least 300 rounds (100 points = 0.01 EUR).
\par
Let us first focus on the two-person route-choice game 
with the payoffs $P_{11} = P_1(2) 
= 0$, $P_{12} = P_1(1) 
= 300$, $P_{21} = P_2(1) 
= -100$, and $P_{22} = P_2(2) 
= -200$ (see Fig.~\ref{fig3}d), corresponding to $C_1 = 600$, $D_1 = 300$, $C_2 = 0$, and $D_2 = 100$.
For this choice of parameters, 
the best individual payoff in each iteration is obtained by choosing route 1 (the ``freeway'')
and have the co-player(s) choose route 2. 
Choosing route 1 is the dominant strategy of the one-shot game, and players 
are tempted to use it. This produces an initial tendency 
towards the ``strongly stable'' user equilibrium \cite{Rapoport} with 0 points for everyone. 
However, this decision behavior is not Pareto efficient in the repeated game. Therefore,
after many iterations, the players often learn to establish the Pareto optimum
of the multi-stage supergame by selecting route 1 in turns (see Fig.~\ref{fig4}). {As a consequence,
the experimental payoff distribution shows a maximum close to 0 points 
in the beginning and a peak at 100 points 
after many iterations (see Fig.~\ref{martin}), which clearly confirms that
the choice behavior of test persons tends to change over time.} Nevertheless, in 7 out of 24 
two-person experiments, persistent cooperation did not emerge during the experiment.
Later on, we will identify reasons for this.

\subsection{Emergence of cooperation and punishment} \label{higher}

\begin{figure}[htbp]
\begin{center}
\includegraphics[width=13cm]{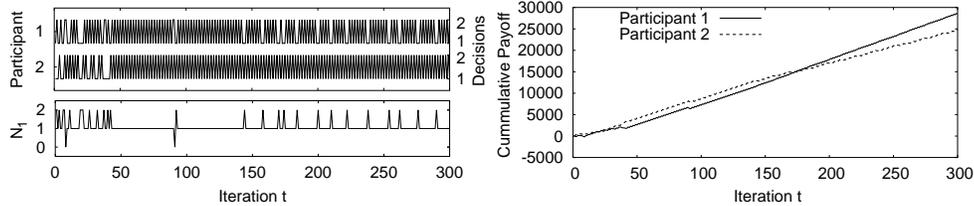} 
\end{center}
\caption[]{Representative example for a 2-person route choice experiment, in which participant 1 
leaves the pattern of oscillatory cooperation temporarily in order to make additional
profits. Note that participant 2 does not ``punish'' this selfish behavior, but continues to take
routes in an alternating way. 
Top left: Decisions of both participants over 300 iterations. Bottom left: Number $N_1(t)$ of 1-decisions
over time $t$. Right: Cumulative payoff of both players as a function of the number of iterations.
The different slopes indicate an unfair outcome despite of high average payoffs of both players.\label{fig7}}
\end{figure}
In order to reach the system optimum of $(-100+300)/2=100$ points
per iteration, one individual has to leave the freeway for one
iteration, which yields a reduced payoff of --100 in favour of a high payoff of +300 
for the other individual. To be profitable also for the first individual, 
the other one should reciprocate this ``offer'' by switching to route 2, 
while the first individual returns to route 1. Establishing this oscillatory cooperative behavior
yields 100 extra points on average. 
If the other individual is not cooperative, both will be back
to the user equilibrium of 0 points only, and the uncooperative individual has temporarily profited
from the offer by the other individual. {This makes ``offers'' for cooperation and, therefore, the 
establishment of the system optimum unlikely.
\par\begin{figure}[htbp]
\begin{center}
\includegraphics[width=8cm]{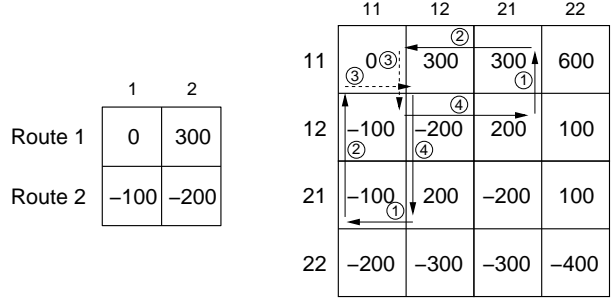} 
\end{center}
\caption[]{Illustration of the concept of higher-order games defined by $n$-stage strategies.
Left: Payoff matrix ${\mathbf{P}} = (P_{ij})$ of the one-shot 2x2 route choice game.
Right: Payoff matrix $\big(P_{(i_1i_2),(j_1j_2)}^{(2)}\big) = ( P_{i_1j_1} + P_{i_2j_2})$ 
of the 2nd-order route choice game defined by 2-stage decisions (right). 
The analysis of the one-shot game (left) predicts that the user equilibrium 
(with both persons choosing route 1) will establish and that no single player could increase the payoff
by another decision. For two-period decisions (right), the system 
optimum (strategy 12 meeting strategy 21) corresponds to a fair solution, but one person can increase
the payoff at the cost of the other (see arrow 1), if the game is repeated. 
A change of the other person's decision can reduce losses and punish this
egoistic behavior  (arrow 2), which is likely to establish the user equilibrium with payoff 0. In order to 
leave this state again in favour of the system optimum, one person will have to make 
an ``offer'' at the cost of a reduced payoff (arrow 3). This offer may be due to a random or intentional
change of decision. If the other person reciprocates the offer (arrow 4), the system optimum is established again.
The time-averaged payoff of this cycle lies below the system optimum.\label{twostage}}
\end{figure}
Hence, the innovation of oscillatory behavior requires intentional or 
random changes (``trial-and-error behavior''). Moreover, the consideration of 
multi-period decisions is helpful. Instead of just 2 one-stage (i.e. one-period) alternative
decisions 1 and 2, there are 
$2^n$ different $n$-stage ($n$-period) decisions. Such multi-stage strategies can be used to
define higher-order games and particular kinds of supergame strategies.
In the two-person 2nd-order route choice game, for example, an encounter of
the two-stage decision 12 with 21 establishes the system optimum and yields equal payoffs
for everyone  (see Fig.~\ref{twostage}). Such an optimal and fair
solution is not possible for one-stage decisions. Yet, the encounter 
of 12 with 21 (``cooperative episode'') is not a Nash equilibrium of the two-stage game, 
as an individual can increase his or her 
own payoff by selecting 11 (see Fig.~\ref{twostage}).
Probably for this reason, the first cooperative episodes in a repeated route choice game
(i.e. encounters of 12-decisions with
21-decisions in two subsequent iterations) do often not persist (see Fig.~\ref{cumulat}). Another
possible reason is that cooperative episodes may be overlooked. This problem, however, can be
reduced by a feedback signal that indicates when the system optimum has been reached.
For example, we have experimented with a green background color. In this setup, a cooperative
episode could be recognized by a green background that appeared in
two successive iterations together with two different payoff values.}
\par\begin{figure}[bp]
\begin{center}
\includegraphics[width=7cm]{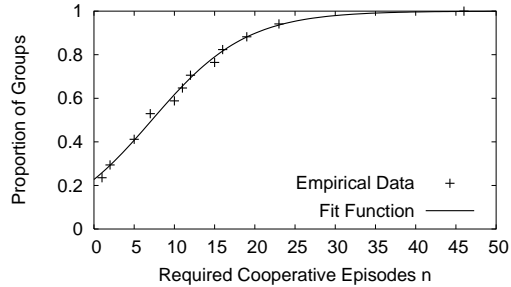} 
\end{center}
\caption[]{Cumulative distribution of required cooperative episodes until persistent
cooperation was established, given that cooperation occured during the duration of the game 
as in 17 out of 24 two-person experiments. The experimental data are well 
approximated by the logistic curve (\ref{slow}) with the fit
parameters $c_2=3.4$ and $d_2=0.17$.\label{cumulat}}
\end{figure}
\begin{figure}[htbp]
\begin{center}
\includegraphics[width=12cm]{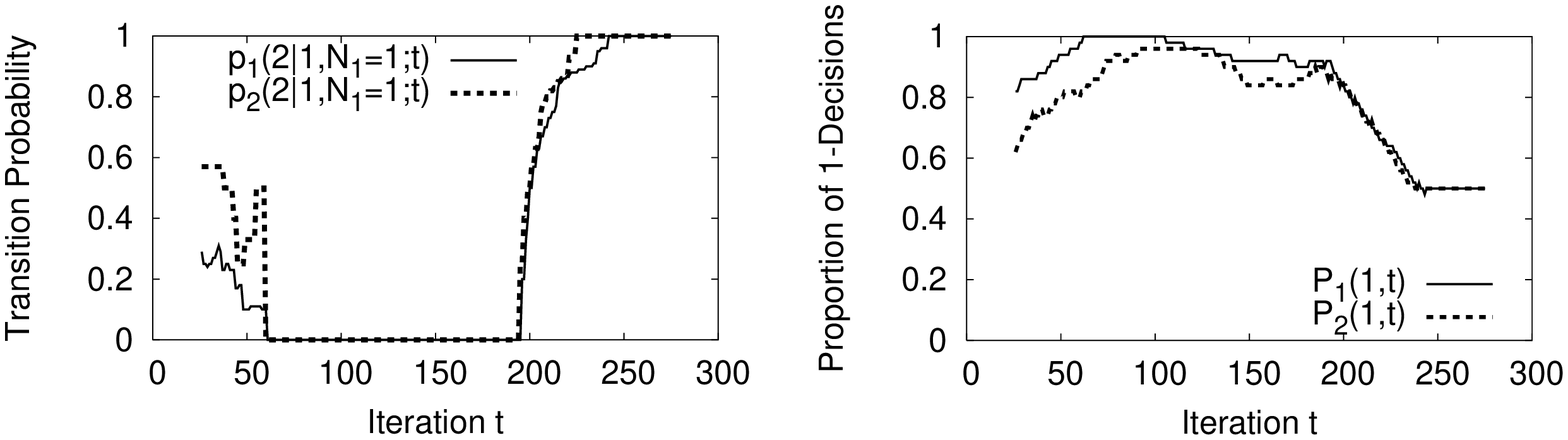} 
\end{center}
\caption[]{Left: Conditional changing probability $p_l(2|1,N_1=1;t)$ of person $l$ from route 1 
(the ``freeway'') to route 2, when the other person has chosen route $2$, averaged over a time
window of 50 iterations. The transition from initially small values 
to 1 (for $t>240$) is characteristic 
and illustrates the learning of cooperative behavior. 
In this particular group (cf. Fig.~\ref{fig4}) the values started even at zero, after a transient time 
period of $t<60$.
Right: Proportion $P_l(1,t)$ of $1$-decisions of both participants $l$ in the 
two-person route choice experiment displayed in Fig.~\ref{fig4}. While the initial proportion is often 
close to 1 (the user equilibrium), it reaches the value 1/2 when persistent oscillatory 
cooperation (the system optimum) is established. 
\label{fig10}}
\end{figure}
\begin{figure}[htbp]
\begin{center}
\includegraphics[width=13cm]{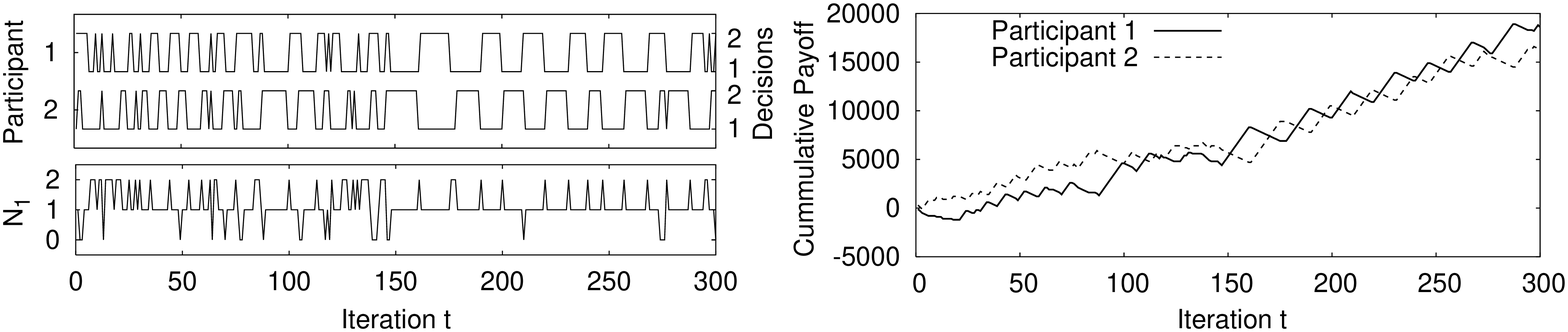} 
\end{center}
\caption[]{Representative example for a 2-person route choice experiment with phase-coordinated oscillations
of long (and varying) time periods larger than 2. 
Top left: Decisions of both participants over 300 iterations. Bottom left: Number $N_1(t)$ of 1-decisions
over time $t$. Right: Cumulative payoff of both players as a function of the number of iterations.
The sawtooth-like increase in the cumulative payoff indicates gains by phase-coordinated alternations 
with long oscillation periods.\label{longer}}
\end{figure}

The strategy of taking route 1 does not only dominate on the first day (in the first iteration).
Even if a cooperative oscillatory behavior has been established, there is a temptation to
leave this state, i.e. to choose route 1 several times, as this yields 
more than 100 points on average for the uncooperative individual at the cost of the participant
continuing an alternating choice behavior
(see Figs.~\ref{fig7} and \ref{twostage}). 
That is, the conditional changing probability $p_l(2|1,N_1=1;t)$ of individuals $l$
from route $1$ to route $2$, when the system optimum in the previous 
iteration was established (i.e. $N_1 = 1$)
tends to be small initially. However, oscillatory cooperation of period 2 needs $p_l(2|1,N_1=1;t) = 1$. 
The required transition in the decision behavior can actually be observed in our experimental
data (see Fig.~\ref{fig10}, left). With this transition, the average frequency of 1-decisions goes down 
to 1/2 (see Fig.~\ref{fig10}, right). Note, however, that alternating reciprocity does not necessarily require oscillations of
period 2. Longer periods are possible as well (see Fig.~\ref{longer}), but have occured only in a few cases
(namely, 3 out of 24 cases).
\par
How does the transition to oscillatory cooperation come about?
The establishment of alternating reciprocity can be supported by a suitable punishment strategy:
If the other player should have selected route 2, but has chosen route 1 instead,
he or she can be punished by changing to route 1 as well, since this causes an average payoff of less than
100 points for the other person (see Fig.~\ref{twostage}). Repeated punishment of uncooperative
behavior can, therefore, reinforce cooperative oscillatory behavior. 
However, the establishment of oscillations also requires costly ``offers'' 
by switching to  route 2, which only pay back in case of alternating reciprocity.
It does not matter whether these ``offers'' are intentional or due to exploratory trial-and-error behavior.
\par
Due to punishment strategies and similar reasons, 
persistent cooperation is often established after a
number $n$ of cooperative episodes. In the 17 of our 24 two-person experiments, in which persistent cooperation was
established, the cumulative distribution of required cooperative episodes could be mathematically described 
by the logistic curve 
\begin{equation}
F_N(n) = 1/[1 + c_N\exp(-d_N n)]
\label{slow}
\end{equation} 
(see Fig.~\ref{cumulat}). Note that, while we expect that this
relationship is generally valid, the fit parameters $c_N$ and $d_N$ 
may depend on factors like the distribution of participant intelligence, 
as oscillatory behavior is apparently difficult to establish (see below).

\subsection{Preconditions for cooperation} 

\begin{figure}[htbp]
\begin{center}
\includegraphics[width=5.2cm]{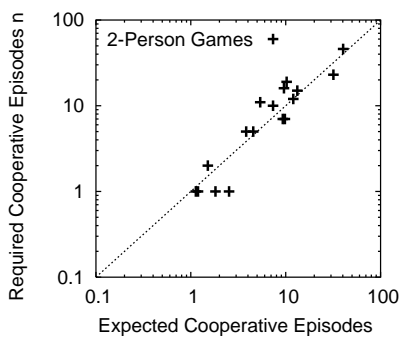} 
\end{center}
\caption[]{Comparison of the required number of cooperative episodes $y$ with the
expected number $x$ of cooperative episodes (approximated as occurence time of persistent
cooperation, divided by the expected time interval $T$ until a cooperative episode
occurs by chance). Note that the data points support the relationship
$y=x$ and, thereby, formula (\ref{coop}).\label{fig12}}
\end{figure}

Let us focus on the time period before persistent oscillatory cooperation
is established and denote the occurence probability that individual $l$
chooses alternative $i\in \{1,2\}$ by $P_l(i)$. The quantity $p_l(j|i)$
shall represent the conditional probability of choosing $j$ in the next iteration, if
$i$ was chosen by person $l$ in the present one. Assuming stationarity
for reasons of simplicity, we expect the relationship
\begin{equation}
p_l(2|1) P_l(1) = p_l(1|2) P_l(2) \, ,
\end{equation}
i.e. the (unconditional) occurence probability $P_l(1,2)=p_l(2|1)P_l(1)$ of having
alternative $1$ in one iteration and $2$ in the next agrees with the joint
occurence probability $P_l(2,1)=p_l(1|2)P_l(2)$ of finding the opposite sequence 21
of decisions: 
\begin{equation}
P_l(1,2) = P_l(2,1) \, .
\label{Eq11}
\end{equation} 
Moreover, if $r_l$ denotes the average changing frequency of person $l$ until persistent
cooperation is established, we have the relation
\begin{equation}
 r_l = P_l(1,2) + P_l(2,1) \, .
\end{equation}
Therefore, the probability that all $N$ players simultaneously change their decision from
one iteration to the next is $\prod_{l=1}^N r_l$. Note that there 
are $2^N$ such realizations of $N$ decision changes 12 or 21,
which have all the same occurence probability because of Eqn.~(\ref{Eq11}). 
Among these, only
the ones where $N/2$ players change from 1 to 2 and the other $N/2$
participants change from 2 to 1 establish cooperative episodes,
given that the system optimum corresponds to an equal distribution over both
alternatives. Considering that the number of different possibilities of selecting $N/2$ out of $N$
persons is given by the binomial coefficient, the occurence probability of cooperative
events is 
\begin{equation}
 P_{\rm  c} 
= \frac{1}{2^N} \left(
\begin{array}{c}
N \\
N/2
\end{array}\right) 
\prod_{l=1}^{N} r_l 
\end{equation}
(at least in the ensemble average). Since the expected time period $T$ until the cooperative state
incidentally occurs equals the inverse of $P_{\rm  c}$, we finally find the
formula
\begin{equation}
 T= \frac{1}{P_{\rm  c}} = 2^N \frac{(N/2)!^2}{N!}
\prod_{l=1}^{N} \frac{1}{r_{l}} \, .
\label{coop}
\end{equation}
This formula is well confirmed by our 2-person experiments (see
Fig.~\ref{fig12}). It gives the lower bound for the expected value of the minimum 
number of required iterations until persistent cooperation can spontaneously emerge
(if already the first cooperative episode is continued forever). 
\par
Obviously, the occurence of oscillatory cooperation is expected to take much longer for a large number $N$
of participants. This tendency is confirmed by our 4-person experiments compared to
our 2-person experiments. It is also in agreement with intuition, as coordination of more people is 
more difficult. (Note that mean first passage or transition times in statistical phyisics
tend to grow exponentially in the number $N$ of particles as well.)
\par
Besides the number $N$ of participants,
another critical factor for the cooperation probability 
are the changing frequencies $r_l$: They are needed for
the exploration of innovative strategies, coordination and cooperation. Although the instruction
of test persons would have allowed them to conclude that 
taking turns would be a good strategy, the changing frequencies $r_l$  of some 
individuals was so small that cooperation within the duration of the respective experiment
did not occur, in accordance with formula (\ref{coop}). The unwillingness of some
individuals to vary their decisions is sometimes called ``conservative'' \cite{press,inSelten,ssrn} or
``inertial behavior'' \cite{Bottazzi}. Note that, if a player never reciprocates ``offers'' by other
players, this may discourage further ``offers'' and reduce the changing frequency of the other 
player(s) as well (see the decisions 50 through 150 of player 2 in Fig.~\ref{fig4}).
\par 
Our experimental time series show that most individuals initially did not know a periodic 
decision behavior would allow them to establish the system optimum. This indicates that
the required depth of strategic reasoning \cite{Colman} and the related
complexity of the game for an average person are already quite high, so that intelligence may matter. 
Compared to control experiments, the hint that the maximum average payoff of 100 points per 
round could be reached ``by variable, situation-dependent decisions'', increased the average 
changing frequency (by 75 percent) and with this the occurence frequency of cooperative events.
Thereby, it also increased the chance that persistent cooperation established during the duration of the 
experiment. 
\par\begin{figure}[bp]
\begin{center}
\includegraphics[width=13cm]{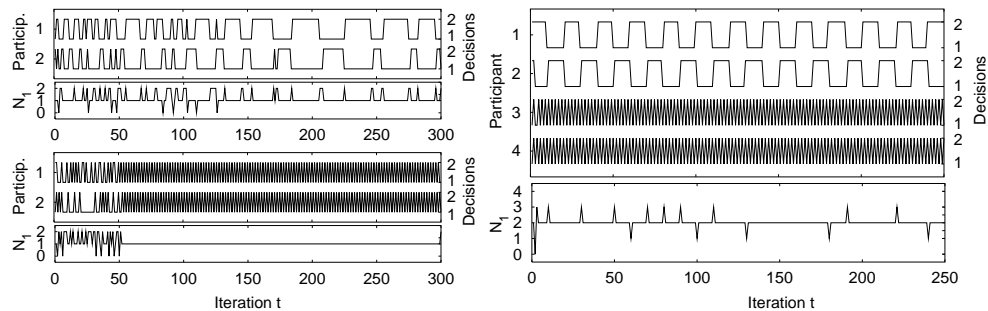} 
\end{center}
\caption[]{Experimentally observed decision behavior when two groups involved in two-person 
route choice experiments 
afterwards played a four-person game with $C_1 = 900$, $D_1 = 300$, $C_2 = 100$, $D_2 = 100$.
Left: While oscillations of period 2 emerged in the second group (bottom), another alternating 
pattern corresponding to $n$-period decisions with $n>2$ emerged in the first group (top).
Right: After all persons had learnt oscillatory cooperative 
behavior, the four-person game just required coordination, but not
the invention of a cooperative strategy. Therefore, persistent cooperation was quickly established
(in contrast to four-person experiments with new participants). It is clearly visible that the test persons 
continued to apply similar decision strategies (right)
as in the previous two-person experiments (left).\label{4person}}
\end{figure}
Note that successful cooperation requires not only coordination \cite{Bottazzi}, but also
innovation: In their first route choice game, 
most test persons discover the oscillatory cooperation 
strategy only by chance in accordance with formula (\ref{coop}). The changing frequency
is, therefore, critical for the establishment of innovative strategies: It determines the exploratory 
trial-and-error behavior. In contrast, cooperation is easy when test persons {\em know} that the 
oscillatory strategy is successful: When two teams, who had successfully cooperated in
2-person games, had afterwards to play a 4-person game, cooperation was {\em always} and quickly 
established (see Fig.~\ref{4person}). In contrast, unexperienced co-players 
suppressed the establishment of oscillatory cooperation in 4-person route choice games.

\subsection{Strategy coefficients}

In order to characterize the strategic behavior of individuals and predict their chances
of cooperation, we have introduced some strategy coefficients. For this, let us introduce the
following quantities, which are determined from the iterations before persistent cooperation is
established:
\begin{itemize}
\item $c_l^k =$ relative frequency of a {\em changed} subsequent decision of individual $l$ if the
payoff was negative ($k=-$), zero ($k=0$), or positive ($k=+$).
\item $s_l^k =$ relative frequency of individual $l$ to {\em stay} with the previous decision if the
payoff was negative ($k=-$), zero ($k=0$), or positive ($k=+$).
\end{itemize}
The Yule-coefficient
\begin{equation}
 Q_l = \frac{c_l^- s_l^+ - c_l^+ s_l^-}{c_l^- s_l^+ + c_l^+ s_l^-}
\end{equation}
with $-1 \le Q_l \le 1$ was used by Schreckenberg, Selten {\em et al.} \cite{inSelten} to
{identify direct responders with $0.5 < Q_l \approx 1$ (who change their decision after a negative
payoff and stay after a positive payoff), and contrarian responders with $-0.5 > Q_l \approx -1$
(who change their decision after a positive payoff and stay after a negative one).} 
A random decision behavior would correspond to a value $Q_l \approx 0$.
However, a problem arises if one of the variables $c_l^-$, $s_l^+$, $c_l^+$, or $s_l^-$ assumes the
value 0. Then, we have $Q_l \in \{-1,1\}$, independently of the other three values. If two of the
variables become zero, $Q_l$ is sometimes even undefined. Moreover, if the
values are small, the resulting conclusion is not reliable. Therefore, we prefer to use the
percentage difference 
\begin{equation}
 S_l = \frac{c_l^-}{c_l^- + s_l^l} - \frac{c_l^+}{c_l^+ + s_l^+}
\end{equation}
for the assessment of strategies. Again, we have $-1 \le S_l \le 1$.
Direct responders correspond to $S_l > 0.25$ and contrarian
responders to $S_l < -0.25$. For $-0.25 \le S_l \le 0.25$, the response to the previous payoff
is rather random.
\par
In addition, we have introduced the $Z$-coefficient
\begin{equation}
 Z_l = \frac{c_l^0}{c_l^0 + s_l^0}  \, , 
\label{Z}
\end{equation}
for which we have $0 \le Z_l \le 1$.
This coefficient describes the likely response of individual $l$ to the user equilibrium. 
$Z_l = 0$ means that individual $l$ does not change routes, if the user equilibrium was reached. 
$Z_l = 1$ implies that person $l$ always changes, while $Z_l \approx 0.5$ indicates a random
response. 
\par\begin{figure}[htbp]
\begin{center}
\includegraphics[width=12cm]{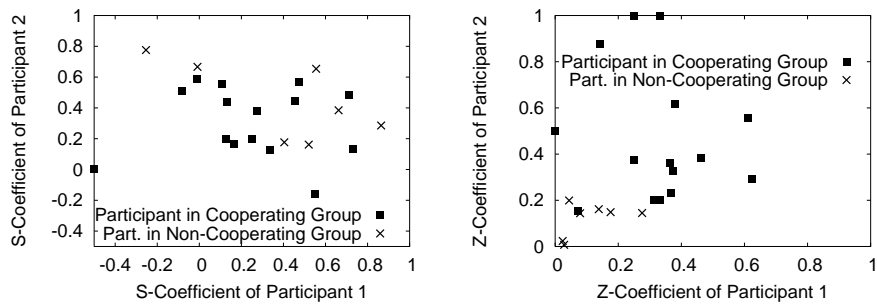} 
\end{center}
\caption[]{Coefficients $S_l$ and $Z_l$ of both participants $l$ in all 24 two-person route choice games.
The values of the $S$-coefficients (i.e. the individual tendencies towards direct or contrarian responses)
are not very significant for the establishment of persistent cooperation, while large enough values of
the $Z$-coefficient stand for the emergence of oscillatory cooperation.\label{phase1}}
\end{figure}
\begin{figure}[bp]
\begin{center}
\includegraphics[width=6.5cm]{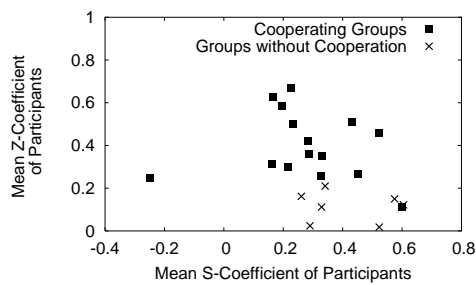} 
\end{center}
\caption[]{$S$- and $Z$-coefficients averaged over 
both participants in all 24 two-person route choice games.
The mainly small, but positive values of $S$ indicate a slight tendency towards direct responses.
However, the $S$-coefficient is barely significant for the emergence of persistent oscillations.
A good indicator for their establishment is a sufficiently large $Z$-value.
\label{phase}}
\end{figure}
Figure~\ref{phase1} shows the result of the 2-person route choice experiments
(cooperation or not) as a function of $S_1$ and $S_2$, and as a function of
$Z_1$ and $Z_2$. Moreover, Figure~\ref{phase} displays the result as a function of
the average strategy coefficients 
\begin{equation}
 Z = \frac{1}{N}\sum_{l=1}^N Z_l
\label{zet} 
\end{equation}
and
\begin{equation}
 S = \frac{1}{N}\sum_{l=1}^N S_l \, .
\end{equation}
Our experimental data indicate that the $Z$-coefficient is a good indicator for the establishment
of cooperation, while the $S$-coefficient seems to be rather insignificant (which also applies to the
Yule coefficient).

\section{Multi-Agent Simulation Model} \label{Sec4}

{In a first attempt, we have tried to reproduce the observed behavior in our 2-person route choice
experiments by game-dynamical equations \cite{Hofbauer}. We have applied these to the 2x2 route choice game
and its corresponding two-, three- and four-stage higher-order games
(see Sec.~\ref{higher}). Instead of describing patterns of alternating cooperation, however,
the game dynamical equations predicted a preference for the dominant strategy of the one-shot
game, i.e. a tendency towards choosing route~1. 
\par
The reason for this becomes understandable through Fig.~\ref{twostage}.
Selecting routes 2 and 1 in an alternating way is not a stable strategy, as the other player can
get a higher payoff by choosing two times route 1 rather than responding with 1 and 2. 
Selecting route 1 all the time even guarantees that the own payoff is never below the one by the other player. However, 
when both players select route 1 and establish the related user equilibrium, no player can
improve his or her payoff in the next iteration by changing the decision.
Nevertheless, it is possible to improve the
long-term outcome, if {\em both} players change their decisions, and if they do it in a coordinated way. 
Note, however, that a strict alternating behavior of period 2 is an optimal strategy only in infinitely 
repeated games, while it is unstable to perturbations in finite games.
\par
It is known that cooperative behavior may be explained by a ``shadow of the future'' \cite{AxeHa81,AxeDi88}, 
but it can also be established by a ``shadow of the past'' \cite{Flache}, i.e. experience-based
learning. This will be the approach of the multi-agent simulation model proposed in this section.
As indicated before, the emergence of phase-coordinated strategic alternation (rather than a statistically 
independent application of mixed strategies) requires an almost deterministic behavior (see Fig.~\ref{repro1}). 
Nevertheless, some weak stochasticity is needed for the establishment of asymmetric cooperation, both for the 
exploration of innovative strategies and for phase coordination. Therefore, we propose the following 
reinforcement learning model, which could be called a 
generalized win-stay, lose-shift strategy \cite{NovakSigmund,Posch}. 
\par\begin{figure}[htbp]
\begin{center}
\includegraphics[width=13cm]{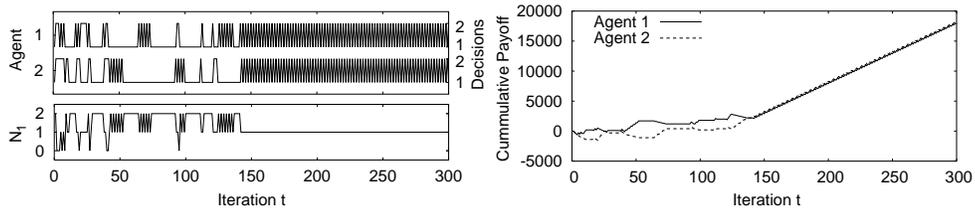} 
\end{center}
\caption[]{Representative example for a 2-person route choice simulation based on 
our proposed multi-agent reinforcement learning model with $P_{\rm av}^{\rm max} = 100$ 
and $P_{\rm av}^{\rm min} = -200$. The parameter $\nu_l^1$ has been set to 0.25. The other
model parameters are specified in the text.
Top left: Decisions of both agents over 300 iterations. Bottom left: Number $N_1(t)$ of 1-decisions
over time $t$. Right: Cumulative payoff of both agents as a function of the number of iterations.
The emergence of oscillatory cooperation is comparable with the experimental data displayed
in Fig.~\ref{fig4}.\label{repro1}}
\end{figure}
Let us presuppose that an individual approximately memorizes or has a good feeling of
how well he or she has performed on average in the last $n_l$ iterations and since 
he or she has last responded with decision $j$ to the situation $(i,N_1)$.
In our success- and history-dependent model of individual decision behavior,
$p_l(j|i,N_1;t)$ denotes agent $l$'s conditional probability of taking decision $j$ at time $t+1$, 
when $i$ was selected at time $t$ and $N_1(t)$ agents had chosen alternative 1. 
Assuming that $p_l$ is either 0 or 1, $p_l(j|i,N_1;t)$ has the meaning of a deterministic
response strategy: $p_l(j|i,N_1;t)=1$ implies that individual $l$ will
respond at time $t+1$ with the decision $j$ to the situation $(i,N_1)$ at time $t$.
\par
Our reinforcement learning strategy can be formulated as follows:
The response strategy $p_l(j|i,N_1,t)$ is switched with 
probability $q_l > 0$, if the average individual payoff since the last comparable situation 
with $i(t') = i(t)$ and $N_1(t') = N_1(t)$ at time $t'<t$ is less than the average 
individual payoff $\overline{P}_l(t)$ during the last $n_l$ iterations. In other words,
if the time-dependent aspiration level $\overline{P}_l(t)$ \cite{Posch,Flache} is not reached by
the agent's average payoff since his  or her last comparable decision,
the individual is assumed to substitute the response strategy  $p_l(j|i,N_1;t)$ by 
\begin{equation}
  p_l(j|i,N_1;t+1) = 1 - p_l(j|i,N_1;t) 
\end{equation}
with probability $q_l$. The replacement of dissatisfactory strategies orients at historical long-term profits (namely, during
the time period $[t',t]$). Thereby, it avoids short-sighted changes after temporary losses. Moreover, it
does not assume a comparison of the performance of the actually applied 
strategy with hypothetical ones as in most evolutionary models. A readiness
for altruistic decisions is also not required, while exploratory behavior (``trial and error'') is
necessary. In order to reflect this, the decision behavior is randomly switched 
from $p_l(j|i,N_1;t+1)$ to $1 - p_l(j|i,N_1;t+1)$ with probability 
\begin{equation}
 \nu_l(t) = \max \left( \nu_l^0, \nu_l^1 
 \, \frac{P_{\rm av}^{\rm max}-\overline{P}_l(t)}
 {P_{\rm av}^{\rm max} - P_{\rm av}^{\rm min}} \right) \ll 1 \, .
\end{equation}
Herein, $P_{\rm av}^{\rm min}$ and $P_{\rm av}^{\rm max}$ denote the minimum and maximum
average payoff of all $N$ agents (simulated players). The parameter $\nu_l^1$ reflects the mutation frequency 
for $\overline{P}_l(t) = P_{\rm av}^{\rm min}$, while the mutation frequency is assumed to be $\nu_l^0 \le \nu_l^1$
when the time-averaged payoff $\overline{P}_l$ reaches the system optimum $\overline{P}_{\rm av}^{\rm max}$.
\par
In our simulations, no emergent cooperation is found for $\nu_l^0=\nu_l^1 = 0$. 
$\nu_l^0 >0$ or odd values of $n_l$ may produce intermittent breakdowns of cooperation. 
A small, but finite value of $\nu_l^1$ is important to find a 
transition to persistent cooperation. Therefore, we have
used the parameter value $\nu_l^1=0.25$, while the
simplest possible specification has been chosen for the other parameters, namely $\nu_l^0 = 0$, 
$q_l = 1$, and $n_l = 2$.}
\par
\begin{figure}[tbp]
\begin{center}
\includegraphics[width=12cm]{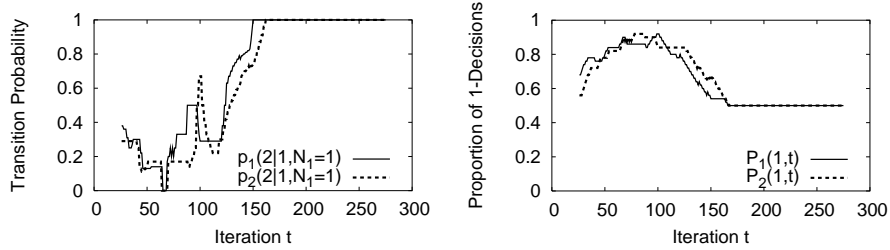}   
\end{center}
\caption[]{Left: Conditional changing probability $p_l(2|1,N_1=1;t)$ of agent $l$ from route 1 
(the ``freeway'') to route 2, when the other agent has chosen route $2$, averaged over a time
window of 50 iterations. The transition from small values to 1 for the computer simulation
displayed in Fig.~\ref{repro1} is characteristic and illustrates the learning of
cooperative behavior. Right: Proportion $P_l(1,t)$ of $1$-decisions of both participants $l$ in the 
two-person route choice experiment displayed in Fig.~\ref{repro1}. While the initial proportion is often 
close to 1 (the user equilibrium), it reaches the value 1/2 when persistent oscillatory 
cooperation (the system optimum) is established. The simulation results 
are compatible with the essential features of the experimental 
data (see, for example, Fig.~\ref{fig10}). \label{repro2}}
\end{figure}

{The initial conditions for the simulation of the route choice game
were specified in accordance with the dominant strategy
of the one-shot game, i.e. $P_l(1,0) = 1$ (everyone tends to choose the freeway initially),
$p_l(2|1,N_1;0)=0$ (it is not attractive to change from the freeway to the side road)
and $p_l(1|2,N_1;0)=1$ (it is tempting to change from the side road to the freeway). Interestingly enough,
agents learnt to acquire the response strategy $p_l(2|1,N_1=1;t)=1$  in the course of
time, which established oscillatory cooperation
with higher profits (see Figs.~\ref{repro1} and \ref{repro2}).}
\par
\begin{figure}[bp]
\begin{center}
\includegraphics[width=12cm]{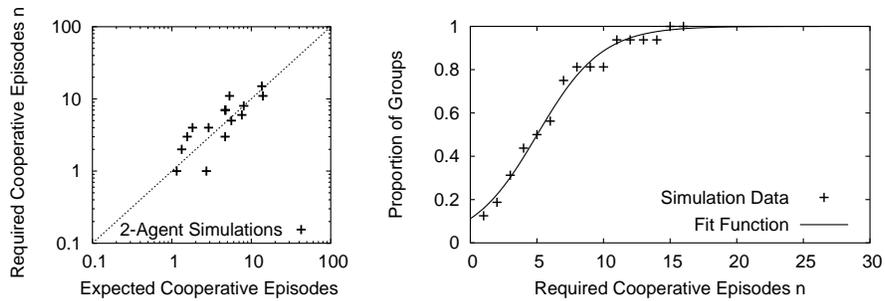}   
\end{center}
\caption[]{Left: Comparison of the required number of cooperative episodes with the
expected number of cooperative episodes in our multi-agent simulation of
decisions in the route choice game. Note that the data points support 
formula (\ref{coop}). Right: Cumulative distribution of required cooperative episodes until persistent
cooperation is established in our 2-person route choice simulations, using the
simplest specification of model parameters (not calibrated). The simulation data are well 
approximated by the logistic curve (\ref{slow}) with the fit
parameters $c_2=7.9$ and $d_2=0.41$.\label{add}}
\end{figure}
\begin{figure}[htbp]
\begin{center}
\includegraphics[width=12cm]{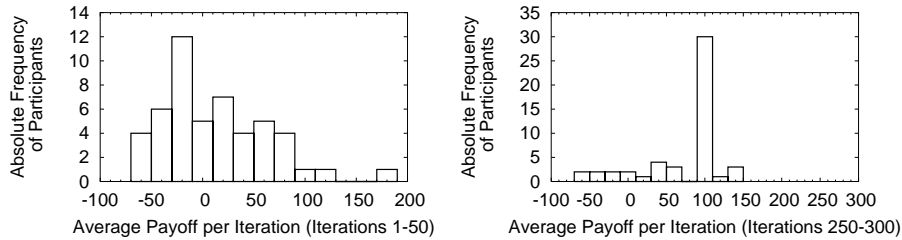} 
\end{center}
\caption[]{Frequency distributions of the average payoffs in our computer simulations of
the 2-person route choice game. Left: Distribution during the first 50 iterations. Right:
Distribution between iterations 250 and 300. Our simulation results are compatible with the
experimental data displayed in Fig.~\ref{martin}.\label{martin2}}
\end{figure}

Note that the above described reinforcement learning model \cite{Flache} responds only to the
own previous experience \cite{Camerer}.  
Despite its simplicity (e.g. the neglection of more powerful, but probably less realistic 
$k$-move memories \cite{BrColman}),
our ``multi-agent'' simulations reproduce the emergence of asymmetric 
reciprocity of two or more players, if an oscillatory strategy of period 2 
can establish the system optimum. This raises the question why 
previous experiments of the $N$-person route choice game \cite{selten,NJP} have observed
a clear tendency towards the Wardrop equilibrium \cite{Wardrop} with $P_1(N_1) = P_2(N_2)$
rather than phase-coordinated oscillations? It turns out that 
the payoff values must be suitably chosen [see Eq.~(\ref{condition})]
and that several hundred repetitions are needed. In fact, the expected time interval $T$ 
until a cooperative episode among $N=N_1+N_2$ participants occurs in our simulations
by chance is well described by formula (\ref{coop}), see Fig.~\ref{add}. The empirically
observed transition in the decision behavior displayed in Fig.~\ref{fig10} is qualitatively
reproduced by our computer simulations as well (see Fig.~\ref{repro2}). The same applies
to the frequency distribution of the average payoff values (compare Fig.~\ref{martin2} with 
Fig.~\ref{martin}) or to the number of expected and required 
cooperative episodes (compare Fig.~\ref{add} with Figs.~\ref{cumulat} and \ref{fig12}).

\subsection{Simultaneous and alternating cooperation in the Prisoner's Dilemma}

\begin{figure}[htbp]
\begin{center}
\includegraphics[width=13cm]{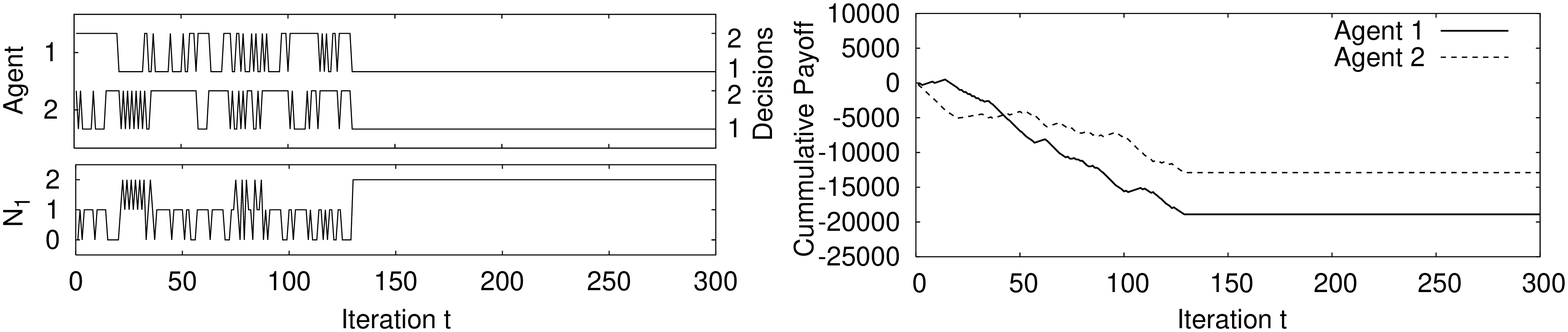} 
\includegraphics[width=13cm]{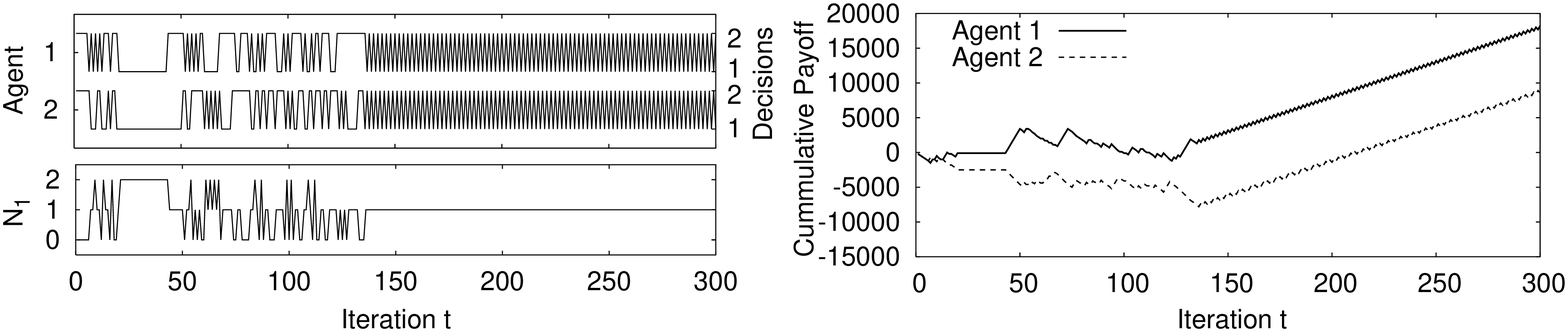} 
\end{center}
\caption[]{Representative examples for computer simulations of
the two different forms of the Prisoner's Dilemma
specified in Fig.~\ref{fig3}b, c. The parameter $\nu_l^1$ has been set to 0.25, while the other
model parameters are specified in the text. 
Top: Emergence of simultaneous, symmetrical cooperation, where
decision 2 corresponds to defection and decision 1 to cooperation. The system optimum
corresponds to $P_{\rm av}^{\rm max} = 0$ payoff points, and the minimum payoff to 
$P_{\rm av}^{\rm min} = -200$. 
Bottom: Emergence of alternating, asymmetric cooperation with $P_{\rm av}^{\rm max} = 100$ 
and $P_{\rm av}^{\rm min} = -200$. Left: Time series of the agents' decisions and the
number $N_1(t)$ of 1-decisions. Right: Cumulative payoffs as a function of time $t$.\label{pris}}
\end{figure}

Let us finally simulate the dynamic behavior in the two different variants of the Prisoner's Dilemma
indicated in Fig.~\ref{fig3}b, c 
with the above experience-based reinforcement learning model. Again, we will assume
$P_{11} = 0$ and $P_{22} = -200$. According to
Eq.~(\ref{condition}), a simultaneous, symmetrical form
of cooperation is expected for $P_{12} = -300$
and $P_{21} = 100$, while an alternating, asymmetric cooperation is expected for
$P_{12} = -300$ and $P_{21} = 500$. Figure~\ref{pris} shows simulation results for the two different cases
of the Prisoner's Dilemma and confirms the two predicted forms of cooperation.
{Again, we varied only the parameter $\nu_l^1$, while we chose the
simplest possible specification of the other parameters $\nu_l^0 = 0$, 
$q_l = 1$, and $n_l = 2$. The initial conditions were specified in accordance with the expected
non-cooperative outcome of the one-shot game, i.e. $P_l(1,0) = 0$ (everyone defects in the beginning),
$p_l(2|2,N_1;0)=0$ (it is tempting to continue defecting), $p_l(1|1,N_1=1;0)=0$
(it is unfavourable to be the only cooperative player), and 
and $p_l(1|1,N_1=2;0)=1$ (it is good to continue cooperating, if the other player cooperates).
In the course of time, agents learn to acquire the response strategy $p_l(2|2,N_1=0;t)=0$ when
simultaneous cooperation evolves, but $p_l(2|2,N_1=1;t)=0$ when alternating cooperation 
is established.}

\section{Summary, Discussion, and Outlook} \label{Sec5}

\begin{figure}[htbp]
\begin{center}
\includegraphics[width=12cm]{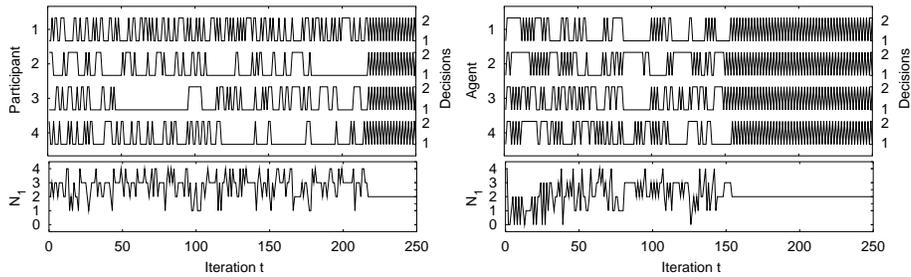} 
\end{center}
\caption[]{Emergence of phase-coordinated oscillatory behavior in the 4-person route choice game
with the parameters specified in Fig.~\ref{4person}.
Left: Experimental data of the decisions of 4 unexperienced participants over 300 iterations. 
Right: Computer simulation with the reinforcment
learning model.\label{four}}
\end{figure}

In this paper, we have investigated the $N$-person day-to-day route-choice game. This special
congestion game has not been thoroughly studied before in the case of small
groups, where the system optimum can considerably differ from the user equilibrium.
The 2-person route choice game gives a meaning to a previously uncommon 
repeated symmetrical 2x2 game and shows a transition from the dominating strategy
of the one-shot game to coherent oscillations, if $P_{12} + P_{21} > 2P_{11}$. However,
a detailed analysis of laboratory experiments with humans reveals that
the establishment of this phase-coordinated alternating reciprocity, which
is expected to occur in other 2x2 games as well, is quite complex. It
needs either strategic experience or the invention of a suitable strategy. Such an innovation is driven 
by the potential gains in the average payoffs of all participants and seems to be  
based on exploratory trial-and-error behavior. If the changing frequency of one
or several players is too low, no cooperation is established in a long time.
Moreover, the emergence of cooperation requires certain kinds of strategies, which can be
characterized by the $Z$-coefficient (\ref{zet}). These strategies can be acquired by means of
reinforcement learning, i.e. by keeping response patterns which have
turned out to be better than average, while worse response patterns are being replaced. 
The punishment of uncooperative behavior can help to enforce
cooperation. Note, however, that punishment in groups of $N>2$ persons 
is difficult, as it is hard to target the uncooperative person, and punishment hits everyone. Nevertheless,
computer simulations and additional experiments indicate 
that oscillatory cooperation can still emerge in route choice games with more than 2 players
after a long time period (rarely within 300 iterations) (see Fig.~\ref{four}). 
\par
Altogether, spontaneous cooperation takes a long time. It is, therefore, sensitive to
changing conditions reflected by time-dependent payoff parameters.
As a consequence, emergent cooperation is unlikely to appear in real traffic systems. This is
the reason why the Wardrop equilibrium tends to occur. However, 
cooperation could be rapidly established by means of advanced traveller information systems (ATIS) 
\cite{bonsall,Hu,MahJou,selten,schreck,Nagel,Johnson,Yama}, which would avoid 
the slow learning process described by Eq.~(\ref{coop}). Moreover, while we
do not recommend conventional congestion charges, a charge for {\em unfair} usage patterns 
would support the compliance with individual route choice recommendations. It
would supplement the inefficient individual punishment mechanism.
\par
Different road pricing schemes have been proposed, 
each of which has its own advantages
and disadvantages or side effects. Congestion charges, for example,
could discourage to take congested routes, which is actually required to reach
minimum {\em average} travel times. Conventional tolls and road pricing may reduce the trip
frequency due to budget constraints, which potentially interferes with economic growth and
fair chances for everyone's mobility.
\par
In order to activate capacity reserves, we therefore
propose an automated route guidance system based on the following principles:
After specification of their destination, drivers should get individual (and, on average, fair)
route choice recommendations in agreement with the traffic situation and
the route choice proportions required to reach the system optimum.
If an individual selects a faster route instead of the recommended route it should use, it will 
have to pay an amount proportional to the decrease in the overall inverse travel time compared
to the system optimum. Moreover, drivers not in a hurry should be encouraged to take the
slower route $i$ by receiving the amount of money corresponding to the related increase in
the overall inverse travel time. Altogether, such an ATIS could support the system optimum while allowing
for some flexibility in route choice. Moreover, the fair usage pattern would be cost-neutral
for everyone, i.e.\ traffic flows of potential economic relevance would not be suppressed by extra costs.
\par
In systems with many similar routing decisions, a 
Pareto optimum characterized by asymmetric alternating cooperation may emerge even
spontaneously. This could help to enhance the routing in data networks \cite{Wolpert} and
generally to resolve Braess-like paradoxes in networks \cite{Cohen}. 
\par
Finally, it cannot be emphasized enough that taking turns is
a promising strategy to distribute scarce resources in a fair and optimal way. It could be applied
to a huge number of real-life situations due to the relevance for many strategical conflicts, including
Leader, the Battle of the Sexes, and variants of Route Choice, Deadlock, Chicken, 
and the Prisoner's Dilemma. The same
applies to their $N$-person generalizations, in particular social dilemmas \cite{socdil1,socdil2,Flache}.
It will also be interesting to find out whether and where metabolic pathways, biological supply networks,
or information flows in neuronal and immune systems use alternating strategies to
avoid the waisting of costly resources.

\section*{Acknowledgements}

D.H. is grateful for the warm hospitality of the Santa Fe
Institute, where the Social Scaling Working Group Meeting in August 2003 
inspired many ideas of this paper. The results shall be presented during the workshop
on ``Collectives Formation and Specialization in Biological and Social Systems'' in Santa Fe
(April 20--22, 2005).

\end{document}